\documentclass[%
superscriptaddress,
unsortedaddress,
 amsmath,amssymb,
 aps,
twocolumn
]{revtex4-2}

\usepackage{graphicx}
\usepackage{dcolumn}
\usepackage{bm}
\usepackage{xcolor,soul}

\usepackage{hyperref}

\usepackage[normalem]{ulem}

\DeclareRobustCommand{\vect}[1]{
  \ifcat#1\relax
    \boldsymbol{#1}
  \else
    \mathbf{#1}
  \fi}

  \newcommand{\av}[1]{\left\langle#1\right\rangle}

\begin{document}

\title{Khinchin's ergodicity and typicality in statistical mechanics}%

\author{Dario Lucente}
\affiliation{Department of Mathematics \& Physics, University of Campania “Luigi Vanvitelli”, Viale Lincoln 5, 81100 Caserta, Italy}
\author{Marco Baldovin}
\affiliation{Institute for Complex Systems, CNR, 00185, Rome, Italy}
\author{Giacomo Gradenigo}
\affiliation{Dipartimento di Fisica e Astronomia “Galileo Galilei”, Università di Padova, Via Marzolo 8, 35131 Padova, Italy}
\author{Angelo Vulpiani}
\affiliation{Institute for Complex Systems, CNR, Universit\`a Sapienza, I-00185, Rome, Italy}
\affiliation{Department of Physics, University of Rome “La Sapienza”, P.le Aldo Moro 5, 00185 Rome, Italy}
\date{\today}
\email{angelo.vulpiani@roma1.infn.it}

\date{\today}

\begin{abstract}
 The formulation of statistical mechanics in terms of ensembles, originally proposed by Gibbs, has proved to be extremely effective in describing the equilibrium state of many-particle systems, in the most diverse contexts (critical phenomena, quantum mechanics, biophysics, just to mention a few). The reasons of this success are still debated, and the connection between dynamics and probability in large systems remains somehow elusive. Indeed, in order to derive the main results of equilibrium statistical mechanics, strong assumptions on the ergodicity of the dynamics are usually required. Nonetheless, empirical observations seem to suggest that the predictions of the theory hold true even when such assumptions are not verified.
 
In this paper we reconsider the point of view put forward by Khinchin, stating that the only relevant ingredients for the validity of statistical mechanics are the large number of degrees of freedom in the system and the choice of extensive observables, irrespectively of the details of the microscopic dynamics. In particular, the presence of dynamical chaos is not required.
To this aim, we discuss some analytical and numerical results on a couple of classical integrable systems, the harmonic chain and the Toda model, showing that many important features predicted by equilibrium statistical mechanics, as for instance Maxwell-Boltzmann distribution, are found even in absence of chaos.
\end{abstract}

\maketitle


\section{Introduction}
\label{sec:intro}

 Establishing a rigorous connection between probability theory and statistical mechanics was one of the most important steps in the development of modern mathematics and physics~\cite{da18,zanghi2005fondamenti}. A primary role in this intellectual challenge was played by ergodic theory, the field of mathematical physics that investigates the statistical properties of deterministic systems. In particular, considerable effort was devoted to understanding under which conditions time averages are well approximated by phase averages, the so-called \textit{ergodic problem}~\cite{eh56,baldovin2025foundations}. 
 
 The abstract formulation of the problem reads as follows.
 Consider a dynamical system
 $(\Omega, S^t,  d\mu(\vect{x}))$, i.e. a deterministic evolution law  in the phase space $\Omega$
\begin{equation}
\vect{x}(0) \to \vect{x}(t)= S^t \vect{x}(0)\,,
\end{equation}
 characterised by the semigroup $S^t: \Omega\to \Omega$
and by a measure $d\mu(\vect{x})$ invariant under the evolution
$S^t$, meaning that $ d\mu(\vect{x})=d\mu(S^{-t}\vect{x})$. 
Given an integrable function $A(\vect{x}): \Omega\to \mathbb{R}$, its time average is defined as
\begin{equation}
\label{eq:tav}
    \overline{A}(\vect{x}_0) \equiv \lim_{{\cal T} \to \infty} {1 \over {\cal T}} 
\int_{t_0}^{t_0+{\cal T}} 
A(\vect{x}(t)) {\rm d} t\,,
\end{equation}
where $\vect{x}_0=\vect{x}(t_0)$ is the initial condition of the system,
while the phase average reads
\begin{equation}
 \langle A \rangle \equiv \int A(\vect{x}) d \mu(\vect{x}) 
  \, .
\end{equation}
The dynamical system 
 is called \textit{ergodic}, with respect to the
measure $d\mu(\vect{x})$, if, for every $A(\vect{x})$ and for almost all choices of $\vect{x}_0$ (with
respect to $\mu$), the equivalence
\begin{equation}
\label{eq:ergostr}
\overline{A}(\vect{x}_0)= \langle A \rangle \, 
\end{equation}
holds. In physics, the most important example is represented by Hamiltonian systems: in this case, $S^t$ is determined
by the Hamilton equations, $\Omega$ is the energy constant hyper-surface  and $ d \mu$ represents
the microcanonical distribution~\cite{ga99}.

From the above definition it is clear that ergodicity, in its strict mathematical sense,  is an extremely demanding property for a mechanical system: it requires time and phase averages to be equal for almost all initial conditions, for every observable $A$. At the same time, it  can happen to be inconclusive at a physical level:
for instance,  the  time  average of some functions may converge to the value predicted by the phase average on time scales that are much larger than the current age of the universe~\cite{ga99}. 

For practical purposes, much weaker forms of ``ergodicity'' are sufficient to assess the validity of statistical mechanics in relevant contexts, as first shown by Khinchin~\cite{kh49}. Two main differences with respect to the strong version are present: (i) the property is required to hold only for the restricted set of observables that can be expressed as sums of functions of the single particle, i.e. those that are relevant for thermodynamics; (ii) it only needs to be verified in the limit of large number of degrees of freedom. Khinchin was able to show that, at equilibrium, these ingredients are sufficient to guarantee the \textit{typicality} of the trajectories in phase space, i.e. the fact that relevant observables do not significantly deviate from their averages. While the validity of ergodicity in its strong sense usually requires the presence of chaos~\cite{da18,gaspard_1998,prigogine1999laws}, recent studies suggest that this weaker form holds also in non chaotic, high dimensional systems~\cite{BGV21,baldovin2025foundations}. In this paper we explicitly discuss the effects of Khinchin's ergodicity in integrable classical systems, focusing in particular on the harmonic chain and Toda model.

The ergodic problem  is a relevant topic also in the domain of  stochastic processes. Here the scenario is usually much simpler than in the  deterministic cases, because, for  many interesting systems, the validity of the ergodic hypothesis is established by rigorous results  (e.g., in Markov chains and Langevin equations~\cite{feller}). However, even for stochastic dynamics, the  ergodic problem can be far from trivial: for instance, in glassy systems characterized by disorder or kinetic constraints, the results may depend on the definition of time average~\cite{benettin2014ergodicity,folena2020rethinking}.

\textcolor{black}{
This paper is organized as follows.
In  Section~\ref{sec:history} we briefly remind the development of the concept of ergodicity
and its role for the statistical physics. Section~\ref{sec:harmonic1} is devoted to the study of the harmonic chain, for which analytical results are available, and the discussion can be more precisely formalized. The distribution of a particle's velocity, in particular, can be shown to tend to the Maxwell-Boltzmann one in the limit of large number of oscillators. In addition, infinitely many observables reaching equipartition are shown to exist and, in general, that equipartition is a {\it typical} property. There, the case of equilibration in disordered chains, where Anderson localization occurs, is also covered. 
In Section~\ref{sec:toda} we discuss some  statistical features of a well known integrable systems,
the Toda model, in the limit of many degrees
of freedom. Contrary to what might be expected from an uncritical interpretation of ergodic theory, numerical simulations provide clear hints that the main prediction of statistical mechanics hold true, despite a patent violation of ``strong'' ergodicity.
In Section~\ref{sec:conclusions} we outline some conclusions and final remarks.}

\section{Ergodicity in statistical physics: from Boltzmann to  simulations}
\label{sec:history}

A detailed  discussion on the history of ergodicity and its relation with probability theory is out of the scope of this paper: the interested reader can refer to~\cite{pl94}. In this section we only recall the most important steps in the development of the concept. 

\subsection{Origins}
The idea of ergodicity  dates back to   Boltzmann's seminal works on statistical mechanics~\cite{da18}, where probabilities are interpreted as frequencies and averages are meant as time integrals. 
Even for experts in the history of  physics
it is not easy  to follow Boltzmann's original ideas. A possible interpretation is given in the celebrated monograph by Paul and Tatiana Ehrenfest~\cite{eh56}:
\begin{quote}
     One
mechanical trajectory fills all of the state space, rendering the time
averages over the trajectory equal to averages over the state space.
\end{quote} 

They  introduced  the distinction between \textit{ergodic}
and \textit{quasi-ergodic} motions; \textcolor{black}{the former are those passing through all points in the phase space}; the latter are those getting arbitrarily
close to any point of phase space (i.e., they are \textit{dense}). The Ehrenfests specified that
the existence of ergodic systems is to be doubted.

The modern  formalization of the  ergodic theory started with the results by Koopman, von
Neumann, and G. D. Birkhoff in the years 1931-1932~\cite{pl94,birkhoff1931proof}. Its development
was very fast, and opened two different research fields: the probabilistic ergodic theory (a branch of measure theory) and the theory
of abstract dynamical systems.  The  fundamental  mathematical achievement by Birkhoff and von Neumann
 was to prove, under physically reasonable assumptions, 
  the following theorems:
  \begin{enumerate}
      \item \label{birk1} For almost all initial conditions $\vect{x}_0$ and every integrable function $A(\vect{x})$, the time average Eq.~\eqref{eq:tav}
 exists.
 \item \label{birk2} The system is ergodic
in the sense of Eq.~\eqref{eq:ergostr} iff the phase space
 $\Omega$ cannot be subdivided into 
two complementary parts that are invariant under $S^t$, both of positive measure (i.e., $\Omega$ is \textit{metrically indecomposable}).
  \end{enumerate}  
Although very important for the field, these results were far from settling the issue. Theorem~\ref{birk1}, at least from a  physical point of view, is
not very restrictive, because the time average Eq.~\eqref{eq:tav} depends in general on the initial condition.  As per Theorem~\ref{birk2}, at a practical level it is only a shift of the problem.

\subsection{Ergodicity and Hamiltonian dynamics}

The study of ergodicity is entangled with an old important problem in Hamiltonian mechanics, i.e. the  existence of conserved quantities other than the total energy, also called \textit{non-trivial first integrals}~\cite{poincare1890probleme}. In order to be ergodic (in the mathematical, or strong, sense), a Hamiltonian system cannot admit any of them: if 
a conserved quantity exists, choosing it as the observable $A(\vect{x})$ would violate Eq.~\eqref{eq:ergostr}, because the left hand side of that equation would depend on the initial condition $\vect{x}_0$, while the right hand side would not.

An extreme case in this sense is represented by \textit{integrable systems}. We recall that a system with  Hamiltonian $ H( {\bf q}, \, {\bf p}) $, where
 $ {\bf q}, \, {\bf p} \in \mathbb{R}^N $,
is called integrable,  if there
exists a canonical transformation  from the original
variables $ ( {\bf q}, \, {\bf p}) $ into a set of new variables
$ ( \vect{I}, \, \vect{\phi}) $,  such that the new Hamiltonian depends
only on $ {\bf I} $:
\begin{equation}
\label{eq:perth}
H[{\bf q}( \vect{I}, \, \vect{\phi}), \, {\bf p}( \vect{I}, \, \vect{\phi})]=H_0 ({\bf I})\, .    
\end{equation}
The new variables are also called \textit{action-angle coordinates}.
If this is the case, there exist $N$ independent first
integrals (the actions $I_n, n=1,...N$), and the motion
evolves on $N$-dimensional tori.  

Let us now consider the perturbed Hamiltonian
\begin{equation}
\label{eq:hpert}
\widetilde{H} ( {\bf I}, \, {\bm \phi})  = H_0 ({\bf I}) + \epsilon 
H_1 ( {\bf I}, \, {\bm \phi})\, ,    
\end{equation}
where $H_1$ is a generic interaction term and $0<\epsilon\ll 1$ is a small coupling. $\widetilde{H}$ is also called \textit{near-integrable} Hamiltonian. We ask whether it still admits non-trivial integrals of motion.
In a  seminal  work on the three body problem, Poincar\'e showed that a  generic near-integrable Hamiltonian,
with $\epsilon \neq 0$, does not allow analytic
first integrals, besides energy. 
This result sounded rather positive for the purposes of statistical mechanics, since it seemed to  suggest that in generic systems, even close to integrable ones, ergodicity holds. In particular, in 1923 the young Fermi generalized Poincar\'e's result, and conjectured that Eq.~\eqref{eq:perth} is always ergodic (unless $\epsilon H_1$ is chosen in such a way that $\widetilde{H}$ is still integrable, which has to be regarded as a pathological case). Following his work, even in the absence of a rigorous
demonstration, the physics community considered the ergodic problem essentially solved~\cite{fermi1923dimostrazione}. 
 In 1930's the ergodic problem lost the interest of the physicists and it continued to be studied only by mathematicians, who tackled it under very general and abstract grounds, disregarding the connections with statistical mechanics.

\subsection{The role of numerical simulations}

After three decades Fermi developed new interest in the ergodic problem. He contributed to its understanding with another important work~\cite{FPU55,ga07}, often referred to as ``FPUT'', after the names of the authors
(Fermi, Pasta and Ulam) and of the programmer of the MANIAC computer who run the simulations (Tsingou)~\footnote{The contribution of M. Tsingou was acknwoledged only recently: this is why the work is sometimes still referred to as ``FPU'' only}.
This work had a
prominent role in the development of 
dynamical chaos and numerical simulations.

Fermi and
collaborators studied numerically the time evolution of $N$ particles of mass
$m$, interacting with non-linear springs, with Hamiltonian
\begin{equation}
\label{fpu}
H= \sum_{i=0}^N \left[  { p_i^2 \over 2 m } + {K \over 2} 
    \bigl( q_{i+1} - q_i \bigr)^2 + {\epsilon \over r} 
 \bigl( q_{i+1} - q_i \bigr)^r
         \right]
\end{equation}
where $q_0=q_{N+1}=0=p_0=p_{N+1}$ and $r=3$ or $r=4$.
For $\epsilon=0$ the system is integrable, as  it is equivalent to
$N$ independent harmonic oscillators
(normal modes) with angular frequencies
\begin{equation}
\label{eq:linear_frequency}
\omega_k = 2\, \sqrt{ {K \over m} } \,  
\sin { k\, \pi  \over 2 (N+1) }  \, ,    
\end{equation}
whose energies  $E_k$ are
constant during the time evolution.
The time average   $\overline{E_k}$ computed along a trajectory
 coincides with 
$ \langle E_k \rangle $ only if $\epsilon \ne 0$, so that the
normal modes interact, loosing memory of their initial conditions. Fermi was interested in understanding what happens if the initial condition is taken
far from equilibrium, e.g. if
all the energy is concentrated in one normal mode:  $E_1 (0) \neq 0$, $E_k (0) = 0 $ for $k=2,\dots N$. Before the FPUT numerical experiment, it was generally expected (based on the
works of Poincar\'e and Fermi himself)  that the first normal mode would
have gradually transferred energy to the others and that, after
some relaxation time, every $E_k(t)$ would fluctuate around the common
value determined by the equilibrium statistical mechanics. 
Instead, no tendency toward equipartition was observed, not even for long times.

A bit paradoxically, the existence of non ergodic behaviour in non-integrable Hamiltonian
systems had already been found by the Soviet mathematician Kolmogorov  one year before the FPUT experiment~\cite{K54,du14}. This fact was surely unknown to the authors. Kolmogorov proposed an important theorem on the existence of regular behaviour for a generic Hamiltonian system of the form Eq.~\eqref{eq:hpert}. The detailed proof was actually completed only later by Arnold  and Moser, but the basic idea was  already clearly expressed in Kolmogorov's work.
The hypotheses of the the theorem are that $H_0 ({\bf I})$ is 
  sufficiently regular, that  $\epsilon$ is small enough and that the relation
  \begin{equation}
  \label{eq:condkam}
      \det \vert \partial^2 H_0 ({\bf I})/
  \partial I_i \partial I_j \vert \neq 0\,,
  \end{equation}
   holds. Actually,  for every value of $\epsilon$ (even very small) some tori of
the perturbed system are destroyed (the so-called \textit{resonant} ones)\textcolor{black}{, and
this forbids analytic first integrals.}  In spite of that, for small
$\epsilon$ most tori survive, although slightly deformed. The
perturbed system's behaviour is therefore akin to the integral, unperturbed one (at least for ``non-pathological'' initial
conditions).

Coming back to FPUT, 
  Izrailev and Chirikov  later noted that for high values of $\epsilon$, when the effects of KAM theorem are switched off, there is a good statistical behaviour. Galgani and his coworkers provided great impulse to the understanding of the role of the dynamics for the validity of the statistical mechanics. The last decades have seen a series of detailed numerical  experiments on the FPUT systems, or similar Hamiltonian, showing rich behaviours that are sometimes very difficult to  disentangle~\cite{benettin2008fermi,campbell2005introduction,galgani1992problem,livi1987chaotic}. Actually, even after seven decades, the problem of thermalization in the FPUT system is still open. A scenario suggested by numerical simulations is the following: the system is ergodic but, if the initial condition is far from thermal equilibrium, the time to reach the equilibrium state diverges when the energy per particle $e\to0$~\cite{benettin2008fermi,BP11}. As a technical remark, we note that for FPUT the ``proper'' $H_0$ is not the Hamiltonian of the harmonic chain but the one of the Toda lattice~\cite{benettin2008fermi,BP11}. 

\subsection{A change of perspective: Khinchin's approach}
Based on the insight coming from KAM theorem and FPUT simulations, one may conclude that Boltzmann's original project to build statistical mechanics on the ergodic hypothesis is hopeless. Actually the program can be still pursued, at the price of weakening the notion of ergodicity. This was shown by Khinchin in his celebrated book~\cite{kh49}, where some important results on the ergodic problem are presented. Khinchin's analysis starts from the following facts:
\begin{enumerate}
    \item in thermodynamic systems the
number  of degrees of freedom $N$ is very large;
\item the relevant observables are not generic
 functions: they present instead a very specific structure, coming from the additivity requirement;
 \item for statistical mechanics to work it is not needed that Eq.~\eqref{eq:ergostr} holds for \textit{all} initial conditions $\vect{x}(0)$: it may be violated in a region of small measure (going to zero as $N \to \infty$).
\end{enumerate}

Kinchin considers a separable Hamiltonian system, i.e.
\begin{equation}
H=\sum_{n=1}^N H_n({\bf q}_n,{\bf p}_n)\,,    
\end{equation}

and a special class of observables (called \textit{sum functions}) of the form
\begin{equation}
f(\vect{x})=\sum_{n=1}^Nf_n({\bf q}_n, {\bf p}_n)\,.    
\end{equation}
Relevant examples of sum functions are given by
pressure, kinetic energy and total energy. Even the empirical single-particle distribution function can be written in that way. The result found by Khinchin is that
\begin{equation}
\textrm{Prob} \left( { {|\overline{f} - \langle f \rangle|} \over
 |\langle f \rangle|} \ge c_1 N^{-1/4} \right) \le c_2 N^{-1/4}
\end{equation}
where $c_1$ and $c_2$ are constants $O(1)$,
and  the probability is computed according to the microcanonical distribution.

Mazur and van der Linden  extended the theorem to systems of particles
interacting through a short range potential~\cite{mazur1963asymptotic}. It is important to stress that in these results the dynamics plays no role, and the existence of good statistical properties is only due to the
fact that $N \gg 1$.



In the following we will show how, even in the absence of chaos, one can have (in
agreement with Khichin's ideas) a good agreement between the time
averages and their values predicted by the equilibrium statistical
mechanics. In the following sections we will always refer to this property as \textit{weak ergodicity}.
We will show that the considered physical dynamics are characterized by this property, by verifying the relation
\begin{equation}
\label{eq:lardev}
    \frac{\av{\overline{f}^2}-\av{\overline{f}}^2}{\av{\overline{f}}^2}\to 0 \quad \text{ for } \quad N\to\infty\,.
\end{equation}
The above condition indicates that in the thermodynamic limit the measure is concentrated around the average, and for a large deviation principle the results holds for almost every initial condition without considering ensemble expectation values. 

\section{Linear Harmonic Systems}
\label{sec:harmonic1}

The weak form of ergodicity discussed above can be tested in systems that are known to violate the strong one. A class of models that are for sure non-chaotic is that of integrable systems.
Due to their amenabilities to analytical computations, let us first focus on harmonic models. Consider a quadratic Hamiltonian $H$
\begin{equation}
\label{eq:Linear-Hamiltonian}
	H= \frac{1}{2}\sum_{i,j=1}^{N} \left[   \frac{p_i^2}{  m_i }\delta_{ij} + K_{ij} q_{i} q_j \right]
\end{equation}
where the masses $m_i$ and the spring stiffnesses are in principle site-dependent. 
In the case of a linear chain with identical constituents (i.e. $m_i=m$ for all $i=1,\cdots,N$), the coupling matrix reads
\begin{equation}
K_{ij} =
\begin{cases}
2K \text{ if } i=j\,;\\
-K \text{ if } i=j\pm1\,;\\
0 \text{ otherwise;}
\end{cases}
\label{eq:linear-chain-coupling}
\end{equation}
and the equation of motion can be easily solved in the Fourier space, leading to the conservation of acoustic modes. More precisely, assuming the chain anchored at both ends to two walls, represented by two virtual fix particles 
$q_0=q_{N+1}=p_0=p_{N+1}=0$, 
with the canonical change of coordinates 
\begin{align}
	P_k &= \sqrt{\frac{2}{N+1}}\sum_{n=1}^{N} p_n \sin\left(\frac{\pi k n}{N+1}\right)\,, \\
	Q_k &= \sqrt{\frac{2}{N+1}}\sum_{n=1}^{N} q_n \sin\left(\frac{\pi k n}{N+1}\right)\,,
    \label{eq:linear-chain-modes}
\end{align}
the Hamiltonian can be rewritten as 
\begin{equation}
\label{eq:Linear-chain-Fourier}
	H= \frac{1}{2}\sum_{k=1}^{N} \left[   \frac{P_k^2}{  m } + \omega_k^2 Q^2_{k} \right]
\end{equation}
where the eigenfrequency $\omega_k$ are those in Eq.~\eqref{eq:linear_frequency}. In the rest of the Section, we discuss how several statistical feature are valid for the harmonic system, even in absence of strong ergodicity. 

\subsection{Maxwell distribution in linear chains}
\label{sec:harmonic2}

A standard way, which is found in textbooks~\cite{peliti2011statistical},  to derive the probability distribution for the velocity, or momentum, 
 of a particle in a system at temperature $T$ is the following.
Consider a system  
with $N$ particles   with  Hamiltonian  
$$
H(\{q_n\} ,  \{p_n\} ) = \sum_{n=1}^N { p_n^2 \over 2m} + \sum_{n, l}V(q_n-q_l) \,.
$$
If the particles are distributed according to the microcanonical distribution
$$
\rho_N(\{q_n\} ,  \{p_n\} ) =  {1 \over \omega(E,N)} \delta(H(\{q_n\} ,  \{p_n\} )  -E) \,,
$$
where
$$
 \omega(E,N)=   \int \int ... \int \delta(H(\{q_n\} ,  \{p_n\} )   -E) \, \prod_{n=1}^N \, dq_n \, dp_n \, ,
 $$
one obtains the single-particle momentum distribution $\rho(p_1)$ by integrating  on $p_2,..., p_N$ and $q_1,..., q_N$:
$$
\rho(p_1)=\int  \int ... \int \rho_N(\{q_n\} ,  \{p_n\} )  \, dp_2 dp_3 ... dp_N   \prod_{n=1}^N \, dq_n \, .
$$
In the limit $N \gg 1$, as originally shown by Maxwell,
$$
\rho(p) \simeq \frac{\sqrt{\beta}}{\sqrt{2 \pi m}} \, e^{ - \beta {p^2 \over 2 m}} \,\,\,\, , \,\,\,\,
\beta= { 1 \over k_B T}= {\partial \ln \omega(E,N) \over  \partial E} \, .
$$
to derive the above  result, which holds for a generic short range potential $V(r)$,
ergodicity in its strong form is assumed to hold. In the case of integrable systems it is therefore necessary to follow  a different   approach.

Consider the linear harmonic chain Eq.~\eqref{eq:linear-chain-coupling} introduced above. $N$ conservation laws hold and, therefore, the system is far from ergodicity in the strong sense. A  derivation of the Maxwell distribution for this system that  is not based on  the ergodic hypothesis has been discussed in~\cite{baldovin2023}.
The idea is to consider the time evolution of the system and
then computing the statistical quantities by performing  an average on a long time interval.
This is nothing but the molecular dynamics method adopted in actual experiments or molecular dynamics simulations. In our case,
due to  the linear structure of the system,
one can write down explicitly  the time evolution of  the $j$-th particle $p_j(t)$
 as a linear combination of the momenta of the normal modes:
$$
p_j(t)=\sum_{k=1}^N c_k  \sin \Big( { j k \pi \over N+1 }  \Big) \sin (\omega_k t + \phi_k)
$$
where the $\{ c_k \}$ and the  $\{ \phi_k \}$ depend on the initial conditions.

We prepare the system by exciting only $N^{\star}\le N$ normal modes, i.e.  $c_k=C$ for $1\le k \le  N^*$  and 
 $c_k=0$  for $ k >  N^*$,  and we choose $\phi_k=0$.
 The assumption $\phi_k=0$, while allowing to make the calculations clearer, is not really
relevant. If $N^{\star}\ll N$ the initial condition is in some sense ``far'' from the typical configurations in thermal equilibrium: it is reminiscent of the numerical experiment in the FPUT paper.
 An explicit computation, see~\cite{baldovin2023}, allows us to compute the average 
 $ \overline{ p^n} $
 of the $n$-th  power of $p$  over a measurement time $T$ 
 $$
\overline{ p^n}={1 \over T} \int_0^T p(t)^n \, dt  \,.
 $$
 At large $T$, one has that the $ \overline{ p^n} $ are 
 zero for odd $n$, while for even values one has
 $$
 \overline{ p^n}= \sigma^{n} (n-1)!!  \,\,\, ,\,\,\ \sigma^2= {\overline{ p^2}}\, .
 $$
 These are  nothing but   the moments of a Guassian distribution, therefore we have that the pdf computed
 from a long time series of $p(t)$ is the  Maxwell-Boltzmann distribution

Let us underline that,  for  the above result, the basic ingredient is the fact that $N$   is very large (and so is $N^{\star}$).
In this sense it can be argued that this result is in agreement with
the general philosophy of Khinchin  on the dominant  role of the many degrees of freedom and the  marginal
relevance of the details of the dynamics for the validity of equilibrium statistical mechanics.
\\
\begin{figure}
    \centering
    \includegraphics[width=0.99\linewidth]{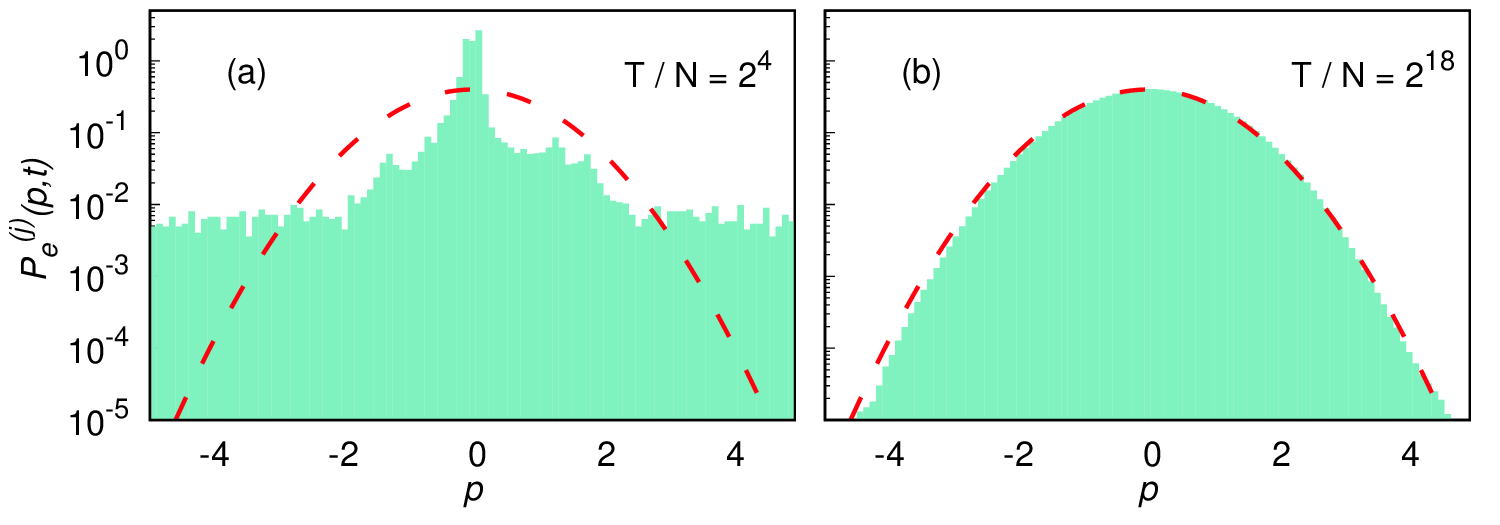}
    \caption{ Empirical density of the momentum of the $j$-th particle, measured in the time interval $[0,T]$, starting from the non-equilibrium initial condition described in the main text. Two values of $T$ are shown. Red curves correspond to the Maxwell-Boltzmann distribution. Left and right panel refer to short and long time evolutions. Here $N=10^3$, $j=123$.}
    \label{fig:Maxwell}
\end{figure}
Fig.~\ref{fig:Maxwell}  shows the histogram of the relative frequencies of the velocities of the $j$-th particle, in the
$N \gg 1$  limit, with the initial conditions discussed before, for two different observation times $T$. As expected from the
computation of the moments  $ \overline{ p^n} $, the shape of the distribution tends to a Gaussian in the
large-time limit.
The fact that the typical time  for the convergence to  equilibrium scales
as $N$ is discussed in~\cite{baldovin2023} and can be understood by
noticing that $\omega_1 \simeq O(1/N)$ and, typically,  $(\omega_{n}-\omega_{n-1})  \simeq 1/N$.

The above result resembles a sort of central limit theorem.  Let us stress, however, that such theorem cannot be directly invoked in this case, because the assumption of weak correlation between the variables does not hold.  This is explicitly shown in~\cite{mazur1960poincare}, where the momentum autocorrelation function $\overline{p_j(t)p_j(t')}$ is computed. Our derivation, instead, relies on Kac's rigorous asymptotic results on  sums of trigonometric functions~\cite{kac1943distribution}
.

\subsection{Thermalization, Equilibration and Typicality of random observables}
Besides the thermal properties of momenta, it has been shown that, in the linear chain and, in general, in harmonic systems, the lack of equilibration property is manifested only when the system is analyzed in the reference frame~\cite{Cocciaglia2022} of the eigenmodes. This can be easily seen by introducing the complex variables 
\begin{align*}
	\psi_k = \frac{P_k+i\omega_k Q_k}{\sqrt{\omega_k}}\,,\\ 
	\psi^{\dagger}_k = \frac{P_k-i\omega_k Q_k}{\sqrt{\omega_k}}\,,\\ 
\end{align*}
such that the Hamiltonian in Eq.~\eqref{eq:Linear-Hamiltonian} takes the form
\begin{equation}
\label{eq:Linear-chain-Fourier-diag}
	H= \frac{1}{2}\sum_{k=1}^{N} \omega_k\psi^\dagger_k\psi_k\,,
\end{equation}
where the energy conservation of each mode is evident. On the other hand, if the system is studied in a reference frame obtained by a random rotation of the eigenvectors $\{\psi_k\}_{k=1,\cdots,N}$, then energy is typically equipartitioned between all the rotated degrees of freedom~\cite{Cocciaglia2022,cattaneo2025thermalization}. To monitor the tendency toward equipartition, one can consider different observables, such as the empirical distribution of the $\mathcal{E}_n$ or the {\it effective number of degrees of freedom} $n_{eff}$.

\begin{figure}
    \centering
    \includegraphics[width=0.99\linewidth]{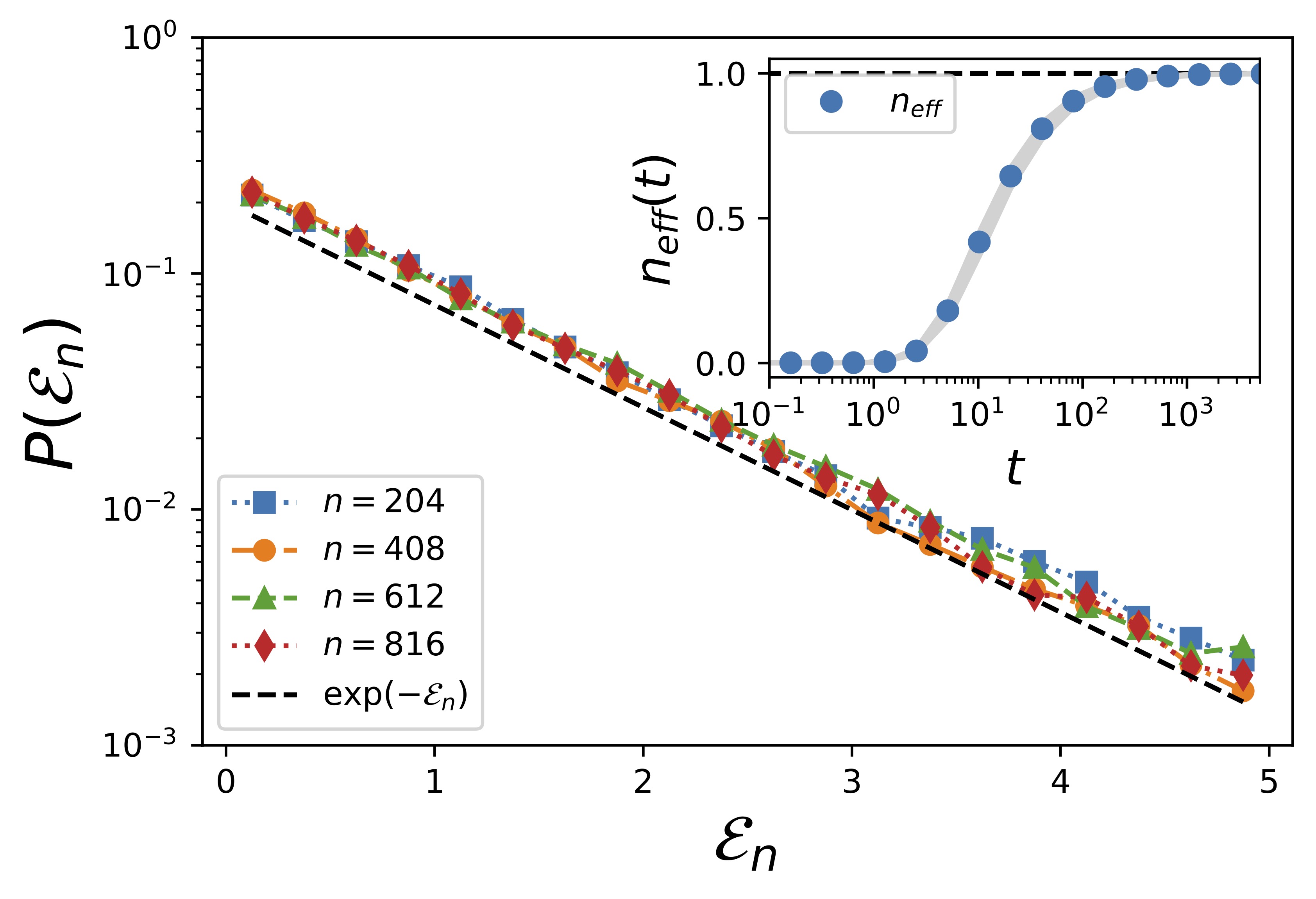}
    \caption{Main panel: probability distribution of the single particle energy of the rotated modes for different beads. The black dashed lines is the theoretical exponential distribution. Inset: effective number of degrees of freedom as a function of time. Blue squares represent an average performed over $30$ realizations of random rotations, while the grey shaded area corresponds to the evolution of each of these $30$ realizations. The horizontal black dashed line corresponds to the equipartition value $n_{eff}=1$.}
    \label{fig:neff}
\end{figure}

In a nutshell, let $\mathcal{R}$ denote a random rotation distributed according to the Haar measure, and $z_n=\sum_k\mathcal{R}_{nk}\psi_k$ the $n-$th rotated mode. In this new reference frame, $H$ reads
\begin{equation}
\label{eq:Linear-chain-rotated}
	H= \frac{1}{2}\sum_{n,n'=1}^{N} \mu_{nn'} z^\dagger_n z_{n'}\,,
\end{equation}
with 
\begin{equation}
\label{eq:mudef}
    \mu_{n n'}=\sum_{k=1}^N \mathcal{R}_{nk}\omega_k\mathcal{R}^\dagger_{k n'}
\end{equation}
and the quantities \begin{equation}
	\mathcal{E}_n(t)=\frac{1}{t}\int_0^t{\rm d}s\,\mu_{nn}z^\dagger_n(s) z_{n}(s)
\end{equation}
are the self-energies of the rotated modes averaged over time.
The empirical distribution at equilibrium is expected to follow an exponential distribution $p(\mathcal{E}_n)\sim \exp{-b \mathcal{E}_n}$, $b$ being related to the mean energy of the system. This is confirmed by the main panel of Fig.~\ref{fig:neff}, where the distribution $p(\mathcal{E}_n)$ is shown for different particles. 
Then, one can define
\begin{equation}
	u_n(t) = \frac{\mathcal{E}_n(t)}{\sum_n \mathcal{E}_n(t)}
\end{equation}
which is the fraction of energy per mode and from that the spectral entropy $S_{sp}(t)$ and the effective number of degrees of freedom $n_{eff}(t)$ as
\begin{align}
	S_{sp}(t)& = -\sum_n u_n \log(u_n)\,,\\
	n_{eff}(t)& = \frac{\exp(S_{sp}(t))}{N}\,.
\end{align}
The quantity $n_{eff}$ is contained in the interval between $0$ and $1$. When it is close to its lower bound, the total energy is shared among a few normal modes; in the opposite limit, the system is close to equipartition. The behaviour of $n_{eff}$, both at the level of single extraction of the rotation matrix $\mathcal{R}$ (grey solid lines) and averaged over $30$ different realizations of the noise (blue circles), is shown in the inset of Fig.~\ref{fig:neff}. 
It can be seen that $n_{eff}$ increases from $0$ to $1$ in a relatively short time for each instance of the random rotation $\mathcal{R}$ proving that the energy is equally shared between the rotated modes $z_k$.  In addition, it can be noted that the averaged behavior does not differ much from the behavior of each instance, suggesting that the equilibration property is typical with respect to the rotation probability distribution. This fact has been shown in~\cite{cattaneo2025thermalization}, where thermalization properties of random rotated modes have been analytically computed. It turned out that the eigenvectors of matrix $\mu$ are {\it delocalized} on the unitary $N-$dimensional sphere implying that equipartition is {\it typical} in the large $N$ limit. In formulae, this insight can be stated as follows. Let $\epsilon_n$ be the relative variance of the time-average observable $\mathcal{E}_n$, i.e.
\begin{equation}
	\epsilon_n = \frac{\langle \mathcal{E}_n^2\rangle-\langle \mathcal{E}_n \rangle^2}{\langle \mathcal{E}_n \rangle^2}
\end{equation}
where the average is taken with respect to the initial condition distribution. Then, using the symbol $\mathbb{E}[\cdot]$ for expressing averages over the rotation distribution, it can be proved that 
\begin{align*}
&\lim_{N\to\infty}\mathbb{E}[\epsilon_n]=0\,,\\
&\lim_{N\to\infty}\frac{\mathbb{E}[\epsilon_n^2]-\mathbb{E}[\epsilon_n]^2}{\mathbb{E}[\epsilon_n]^2}=0\,.
\end{align*}
This means that for almost all rotations $\mathcal{R}$ distributed according to the Haar measure the energy of rotated modes reaches equipartition while the number of rotations which do not lead to it vanishes as the number of degrees of freedom $N$ tends to infinity. Interestingly, whether the asymptotic value of $\mathcal{E}_n$ agree with the predictions of microcanonical ensemble depends upon the initial condition distribution~\cite{cattaneo2025thermalization}. Clearly, if the initial condition are drawn from the microcanonical distribution, then $\lim_{t\to\infty}\mathcal{E}_n(t)$ coincides with the microcanonical average, meaning that the system not only {\it equilibrates}, but also {\it thermalizes}. Moreover, it is possible to prove that thermalization is typical even if only a fraction $N_s$ of the normal modes is excited at the beginning, provided that some suitable assumptions about the initial distribution are met~\cite{cattaneo2025thermalization}. On the other hand, when these conditions are not satisfied the system does not thermalize and microcanonical averages differ from temporal averages. Note that the results in~\cite{cattaneo2025thermalization} on the equipartition of rotated modes hold independently of the specific form of the dispersion relation $\omega_k$ and of the eigenvectors $\psi_k$.  Moreover, they can be generalized to every linear Hamiltonian in the form of Eq.~\eqref{eq:Linear-Hamiltonian} (which can be easily recast as Eq.~\eqref{eq:Linear-chain-rotated}), whose eigenvectors are delocalized over the unit-sphere. The above discussion can be seen as an example of the large deviation principle Eq.~\eqref{eq:lardev} stated at the end of Sec.~\ref{sec:history}. The ensemble averages over the microcanonical distribution and the averages over the rotation distribution (i.e. the Haar measure) play the role of the expectation values in Eq.~\eqref{eq:lardev}.


\subsection{Thermalization and Localization in Harmonic Chains}
Since most of the results presented in the previous section rely on the notion of delocalization of eigenstates of the matrix $\mu$ defined by Eq.~\eqref{eq:mudef}, it is natural to discuss the case where these eigenvectors are localized. It is well known that the linear chain with random couplings (or random masses) displays localization, meaning that each eigenvector $\psi_k$ decays exponentially fast from its maximum, with a characteristic length $\xi_k$ related to the disorder distribution~\cite{matsuda1970localization,ishii1973localization,thouless1974electrons,crisanti2012products}. Without entering into details of localization, 
 here we want to discuss its relation to statistical mechanics and typicality of classical systems.
On the one hand, the exponential localization of eigenvectors strongly impacts transport phenomena in these systems (see~\cite{matsuda1970localization,ishii1973localization,sierant2025many} and references therein).
 From the general discussions above, it should be clear that when the eigenstates of the matrix $\mu$ are strongly localized, there is no reason to expect the observables $\mathcal{E}_n$
 to thermalize. Indeed, in this regime, each observable $\mathcal{E}_n$ depends on a {\it few} integral of motions only, where the precise number depends on the localization lengths $\{\xi_k\}_{k=1,\cdots,N}$, and, therefore, these observables cannot be written as a superposition of many functions depending on an extensive number of conserved quantities, hence violating one of Khinchin's conditions. 
  On the other hand, all results derived for rotated modes apply also in this situation, meaning that infinitely many observables with good thermalization properties can be designed. We emphasize that those results hold at the level of single realizations for the majority of the initial conditions, as long as the rotated degrees of freedom are concerned. This is because, in the reference frame of the rotated modes, the eigenfunctions are delocalized (for instance, a delta function in real space is a constant in Fourier representation). In this sense, the lack of thermalization in systems displaying localization has to be ascribed to the choice of the local observables in real space rather than to an intrinsic peculiarity of these models. 
 To support this point of view, 
we consider a linear chain where the spring stiffnesses are random variables uniformly distributed. We focus on
{\it macroscopic} observables in the position reference frame, i.e. those which are written as sums over many sites $n$ of functions $f_n(q_n)$.

\begin{figure}
    \centering
    \includegraphics[width=0.99\linewidth]{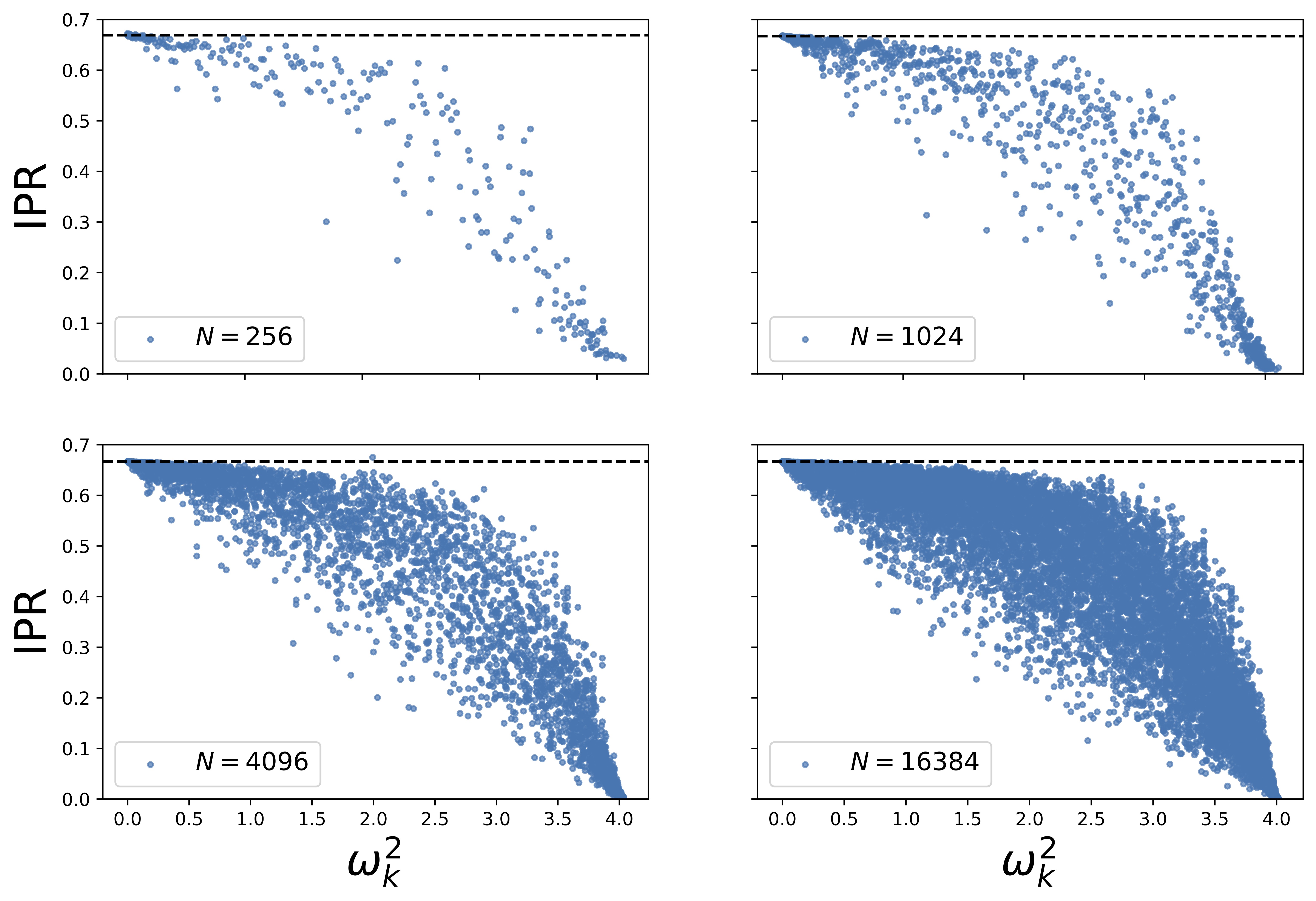}
    \caption{Values of $\text{IPR}$ as function of the eigenfrequency $\omega_k^2$. Different panels corresponds to different system sizes $N$. The dashed black lines is the value of the $\text{IPR}$ of the linear chain without disorder, when the eigenvectors are all delocalized.}
    \label{fig:disorder_spectrum}
\end{figure}

\begin{figure}
    \centering
    \includegraphics[width=0.99\linewidth]{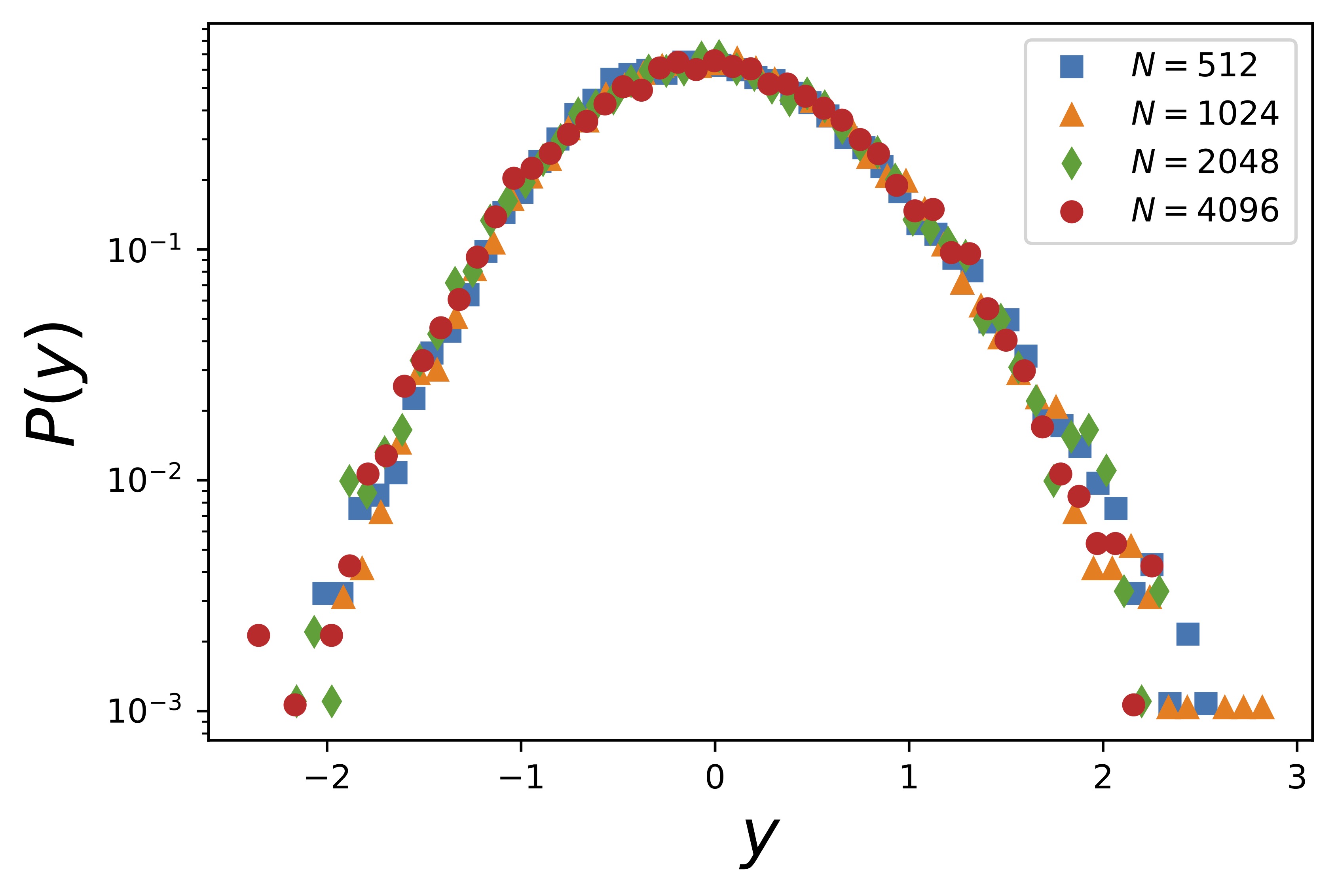}
    \caption{Distributions of energy per particle $e_c$ computed on a subsystem of size $N_c=\frac{N}{4}$ and rescaled by the system size as in Eq.~\eqref{eq:y}. Different symbols refer to different values of $N$. }
    \label{fig:thermalization_disorder}
\end{figure}

In order to compare systems with different number of beads $N$, the variance of the disorder $\sigma^2$ has been chosen equal to $\sigma^2=N^{-1}$. In this way, the demarcation frequency $\omega_d$ between localized and non-localized eigenvectors does not change in the thermodynamic limit~\cite{ishii1973localization}. This can be appreciated considering the {\it inverse participation ratio} (IPR) of each eigenvector, which for a generic vector $v$ reads 
\begin{equation}
    \text{IPR} = \frac{\sum_{i}v_i^2}{N \sum_{i}v_i^4}\,,
\end{equation} 
and is a measure of the localization of the vector $v$: $\text{IPR}=1$ in the extreme case of a perfect delocalized vector $v_i=\frac{1}{\sqrt{N}}$ for all $i$, while $\text{IPR}=\frac{1}{N}$ when $v_i=\delta_{ij}$, i.e. when only a single component of $v_i$ is different from $0$. 
Fig.~\ref{fig:disorder_spectrum} shows the $\text{IPR}$ of each eigenvector of a disordered chain as a function of the eigenfrequencies $\omega_k^2$. When the number of particles $N$ increases, the distribution of the values of $\text{IPR}$ becomes broader, but the average shape remains almost the same. Note that the asymptotic value for $\omega_k\to0$ corresponds to the $\text{IPR}$ of the linear chain, (Eq.~\eqref{eq:linear-chain-modes}), i.e.  
$\text{IPR}=\frac{2}{3}(1+N^{-1})$. 
As a macroscopic observable we consider the infinite time average of the energy $E_c$ of a subsystem of length $N_c$, namely
\begin{equation}
    \bar{E_c} = \lim_{T\to\infty }\frac{1}{T}\int_0^{T} {\rm d}t\,E_c(t)\,,
\end{equation}
where
\begin{equation}
    E_c = \sum_{i=n_0+1}^{n_0+N_c} \left[   \frac{p_i^2}{ 2 m } + \frac{K_{i+1}}{4} (q_{i+1} -q_i)^2 + \frac{K_{i}}{4}(q_{i} -q_{i-1})^2\right]\,
\end{equation}
with $1\le N_c \le N$ and $1 \le n_0 \le N-N_c$.
In the limit $N\to\infty$, $\frac{N_c}{N}\to const$, if the system was ergodic the quantity $\bar{E_c}$ would be distributed around $N_c e$, being $e$ the single particle energy $e=\frac{E_{tot}}{N}$, with standard deviation $\sigma_{E_c}\sim O(\sqrt{N_c})$. To check whether this property holds even in the absence of strong ergodicity, we consider the distribution of  the intensive quantity 
\begin{equation} 
\label{eq:y}
y = \left(\frac{N}{E_{tot}N_c}\bar{E_c}-1\right)\sqrt{N_c}\,.
\end{equation}
 The initial conditions are drawn from a ``restricted'' microcanonical ensemble, where only the momenta of $N_c$ sites are excited. These momenta $P_n$ are normally distributed and rescaled in order to fix the total energy $E_{tot}=\frac{N}{2}$. Fig.~\ref{fig:thermalization_disorder} shows the distributions of $y$ for $N_c=\frac{N}{4}$ obtained from $10^5$ different initial conditions. As it can be noted, the distributions corresponding to different $N$ collapse onto a single curve, meaning that the relative fluctuations of $\bar{E_c}$ vanish in the $N\gg1$ limit and typicality holds. Let us stress once again that the initial conditions we consider are not typical with respect to the microcanonical ensemble, and therefore show that also the overwhelming majority of zero-measure initial conditions lead to results in agreement with statistical mechanics.   
Thus, from our perspective, localized systems perfectly fit in the scenario described in this work, where the validity of statistical mechanics basically relies on the large number of constituents of the system and the good thermal behavior of macroscopic observables. 

 We note, however, that, for some systems, localization plays a major role in explaining the physical properties observed in real space. For instance, the insulating properties of disordered crystals can be explained taking into account that the localization of eigenmodes prevents electrons from freely diffuse~\cite{anderson1958,thouless1974electrons}. Similar behavior can be observed for heat transport phenomena where localization leads to a vanishing thermal conductivity~\cite{matsuda1970localization,ishii1973localization,dhar2008heat}, or in quantum disordered system showing many body localization~\cite{sierant2025many}. 
Although this phenomenology may appear in contrast with the weak ergodicity presented here, it should be noted that in these cases, the initial conditions are not typical with respect to the microcanonical distribution. In this sense, they belong to set of zero measure that is discarded in Khinchin's approach.

\section{Toda model}
\label{sec:toda}
\begin{figure}
    \centering
    \includegraphics[width=0.99\linewidth]{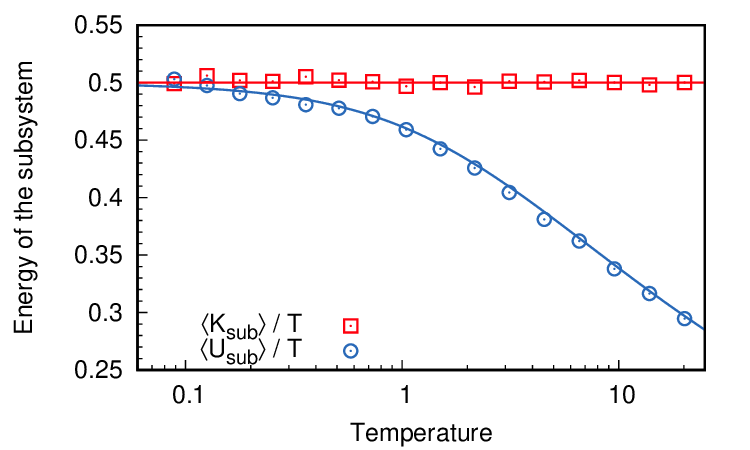}
    \caption{Energy per particle in a subsystem. Red squares and blue circle refer to kinetic and potential energy respectively, normalized with temperature. The solid lines correspond to the theoretical predictions. Here $N=256$, $N_c=16$.}
    \label{fig:thermo}
\end{figure}

\begin{figure}
    \centering
    \includegraphics[width=0.99\linewidth]{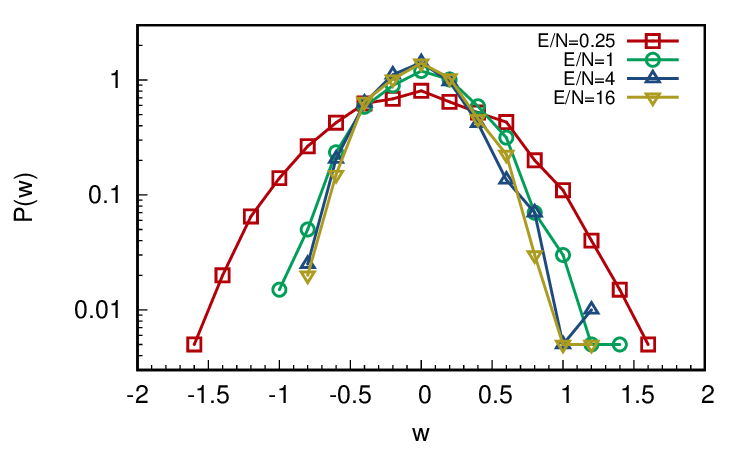}
    \caption{Distribution of the quantity $w(t^{\star})$ defined in Eq.~\eqref{eq:w}, for different values of the energy. The histograms are obtained by taking many different initial distributions and measuring the value of the energy after a long time ($t^{\star}=1000$). Here $N=256$, $N_c=64$.}
    \label{fig:toda_distr}
\end{figure}

While the previous discussions are focused on linear systems, 
 in the present section we consider the more involved case of non-linear integrable systems.  In particular, we study the Toda model, introduced in~\cite{toda1970waves} -- representing a system of $N$ particles connected by non-linear springs, whose Hamiltonian reads
\begin{equation}
\label{eq:Toda-Hamiltonian}
H= \sum_{i=1}^{N} \left[  { p_i^2 \over 2  } + V\bigl( q_{i+1} - q_i \bigr)\right]
\end{equation}
with $V(x)=e^{-x}+x-1$.
This model played a major role both for understanding differential equation admitting infinitely many local conserved-quantity~\cite{toda1975studies} as well as for understanding the slow relaxation toward thermal state in FPUT being its nearest integrable system~\cite{BP11}. The integrability of the Toda model was first demonstrated by Henon, who explicitly wrote down the expression of $N$ conserved quantities in involution between themselves~\cite{henon1974integrals}. By employing a technique developed by Lax in the context of partial differential equations~\cite{lax1968integrals,Lepri2025}, Flaschka exhibited a different pathways for deriving the conservation law in this model~\cite{flaschka1974toda}. Consider a system of $N$ particles with periodic boundary condition $q_{N+1}=q_1$ with a symplectic evolution ruled by the Hamiltonian in Eq.~\eqref{eq:Toda-Hamiltonian}. Let us introduce the variables
\begin{equation}
	a_n =\frac{1}{2} e^{-\frac{q_{n+1}-q_{n}}{2}} \,; \qquad b_n = -\frac{1}{2}p_n\,.
\end{equation}
The equations of motion for $a_n$ and $b_n$ read
\begin{align}
&\dot{a}_n =  a_n(b_{n+1}-b_n)
\nonumber\\
&\dot{b}_n = 2(a_{n+1}^2-a_n^2)
\label{eq:evolution_ab}
\end{align}
Defining the two matrices $L$ and $B$ as 
\begin{align}
&L_{i,j}=\begin{cases}
& a_i \text{ if } i=j\,;\\
& b_i \text{ if } j=i-1 \text{ or } j=i+1\,;\\
& 0 \text{ otherwise}\,;\\
\end{cases}
\nonumber\\
&B_{i,j}=\begin{cases}
&b_i \text{ if } j=i+1\,;\\
&-b_i \text{ if } j=i-1\,;\\
& 0 \text{ otherwise}\,;
\end{cases}
\end{align}
and using Eq.~\eqref{eq:evolution_ab}, it can be shown that the following relation holds~\cite{flaschka1974toda}
\begin{equation}
	\frac{d}{dt}L = [L,B]\,.
	\label{eq:derivative_Lax}
\end{equation}
Moreover, Eq.~\eqref{eq:derivative_Lax} implies that the eigenvalues ${\lambda_n}$ of $L$ are constant~\cite{flaschka1974toda}. {\color{black} This means that also the coefficients of characteristic polynomial of $L$ are independent of time, and are equivalent to the integrals of motion founded by Henon~\cite{henon1974integrals}.} 
In analogy with conservation laws of local operators emerging in the context of wave equations, we can consider as integral of motion ${I_n}$ the trace of the $n-$th power of $L$, that is 
\begin{equation}
	\mathcal{I}_n = \text{Tr}(L^n)=\sum_{i=1}^N\lambda_i^n\,.
\end{equation}
{\color{black} These conserved quantities have been recently used for providing a description in terms of the Generalized Gibbs Ensemble of the Toda model~\cite{spohn2020generalized}. As will be apparent from the following discussion, such a description is needed to properly account the hydrodynamics of the system~\cite{spohn2020generalized}, but it is not necessary as long as the equilibrium properties of certain observables are considered. 
 Indeed, in~\cite{BGV21} it has been shown that for sufficiently high temperatures the energies of normal modes reach equipartition despite the integrability of the model. This fact does not contradict the strong non-equipartion displayed at extreme low energy, related to the states appearing in FPUT experiments~\cite{BP11}. Indeed, at very low energy, the Toda modes are not collective functions of the Fourier modes, therefore Khinchin's condition on the extensivity of the observable is violated.} On the same line of~\cite{BGV21}, we will show that statistical mechanics predictions hold despite the lack of ergodicity of the system. To this end, let us consider the canonical partition function $\mathcal{Z}_N$ of the Toda model
\begin{align}
	\mathcal{Z}_N &= \int \prod_{i=1}^N{\rm d}p_i \,{\rm d}q_i e^{-\beta\sum_{n=1}^{N} \left[  { p_n^2 \over 2  } + V\bigl( q_{n+1} - q_n \bigr)\right]}\nonumber\\
	& = \left(\frac{2\pi }{\beta}\right)^{\frac{N}{2}}\int \prod_{i=1}^N{\rm d}q_i e^{-\beta\sum_{n=1}^{N}V\bigl( q_{n+1} - q_n \bigr)}\nonumber\\
	& =  \left(\frac{2\pi }{\beta}\right)^{\frac{N}{2}}\mathcal{Z}^{(V)}_N
\end{align}
where $\mathcal{Z}^{(V)}_N$ is the part related to the potential energy $V$, which can be computed using a large deviation principle as outlined below.
Performing the change of variables $q_{cm}=\frac{1}{N}\sum_n q_{n}$, $r_n=q_{n+1}-q_{n}$ for $i=1,\cdots, N$ with $\sum_n r_n=0$,  and considering that, since the system is translationally invariant, we can impose $q_{cm}\equiv 0$ without loss of generality, $\mathcal{Z}^{(V)}_N$ can be rewritten as
\begin{align}
	\mathcal{Z}^{(V)}_N &= 
	\int \prod_{i=1}^{N}{\rm d}r_i e^{-\beta\sum_{n=1}^{N}V\left( r_{n} \right)} \delta\left(\sum_{l=1}^N r_l \right)\nonumber\\
	& =  \int_{s_0-i\infty}^{s_0+i\infty}\frac{{\rm d}s}{2\pi i} \int \prod_{i=1}^{N}{\rm d}r_i e^{-\beta\sum_{n=1}^{N}\left[V\left( r_{n} \right)-\frac{s r_n}{\beta}\right]}\nonumber\\
	& = \int_{s_0-i\infty}^{s_0+i\infty}\frac{{\rm d}s}{2\pi i} \left[\int {\rm d}r e^{-\left[\beta V\left( r \right)-s r\right]}\right]^N\nonumber\\
	& = e^{N\beta}\int_{s_0-i\infty}^{s_0+i\infty}\frac{{\rm d}s}{2\pi i}\,  \left[z_\beta(s)\right]^N\,,
\end{align}
having defined 
\begin{align}
	z_\beta(s) &= \int {\rm d}r \,e^{-\left[\beta e^{- r}-s r\right]}\nonumber\\
	& = \int \frac{{\rm d}t}{t}\,e^{-\left[t +s \log\left(\frac{t}{\beta}\right)\right]}\nonumber\\
	& = \beta^{s}\int \frac{{\rm d}t}{t^{s+1}}\,e^{-t}=\beta^{s}\Gamma(-s)\,,
\end{align}
where we have considered the change of variable $t=\beta e^{-r}$.
In the limit $N \gg 1$, the partition function is asymptotically equivalent to
\begin{align}
	\mathcal{Z}^{(V)}_N & = e^{N\beta}\int_{s_0-i\infty}^{s_0+i\infty}{\rm d}s\,  \beta^{N s}\Gamma(-s)^N \sim e^{N\beta} \beta^{N s^\star}\Gamma(-s^\star)^N
	\label{eq:partition_function_potential}
\end{align}
being $s^\star(\beta)=\inf_s\{s\log(\beta)+\log(\Gamma(-s))\}$. From Eq.~\eqref{eq:partition_function_potential}, we can derive the expression of the average potential energy $\langle u \rangle$ as
\begin{align}
	\langle u \rangle = -\frac{1}{N}\partial_\beta\log\mathcal{Z}^{(V)}_N & = -(1+\frac{s^\star}{\beta})\,.
	\label{eq:average_potential_energy}
\end{align}
The behavior of $\langle u \rangle$ as a function of $\beta$, requires the explicit solution of the problem $\log(\beta)=-\frac{\Gamma'(-s^\star)}{\Gamma(-s^\star)}$ which is not known in general. What can instead be predicted are the asymptotic limit $\beta\to\infty$ and $\beta\to0$, yelding
\begin{align}
\langle u \rangle &= \frac{1}{2\beta} \text{ for } \beta\to\infty\\
\langle u \rangle &=-\frac{1}{\beta\log(\beta)} \text{ for } \beta\to0.
\end{align} 
This theoretical picture, as well as Khinchin's ideas, can be tested by performing numerical simulations. The numerical experiment we carried out consists in assigning an initial condition in such a way that the momenta are distributed according to a Maxwell-Boltzmann distribution while the positions $q_n$ are all $0$. To guarantee that the momentum of the center of mass vanishes exactly, we explicitly subtract it from the initial value of each momentum. 
The first observable we consider is the ``local'' temperature of a small subchain of length $N_c$, i.e. the long-time average of kinetic energy per degree of freedom 
\begin{equation}
K_{sub}=\frac{1}{2N_\star}\sum_{i=k+1}^{k+N_\star} \frac{1}{\mathcal{T}}\int_0^\mathcal{T}{\rm d}t\, p_i^2(t)\,.
\end{equation}
From the theory, we expect such a quantity to be almost constant and equal to half the temperature $T=\frac{1}{N}\sum_{i=1}^{N} p_i^2(0)$. Fig.~\ref{fig:thermo}, shows that such a prediction holds in the whole range of considered $T$. We are not considering any phase space average and each point has been computed from a single long trajectory. For comparison with Eq.~\eqref{eq:average_potential_energy}, we also computed the long-time average of the potential energy for the small subchain, that is
$$\bar{u}=\frac{1}{N_\star}\frac{1}{\mathcal{T}}\int_{0}^{\mathcal{T}} U_{sub}(\mathbf{q})$$
where $U_{sub}(\mathbf{q})= \sum_{i=k+1}^{k+N_\star} V\bigl( q_{i+1} - q_{i} \bigr)$. Fig.~\ref{fig:thermo} reports this quantity together with the analytical prediction (Eq.~\eqref{eq:average_potential_energy}). 
{\color{black} Concerning the long-time average of the internal energy, it clearly shows that the canonical ensemble (with a single inverse temperature $\beta$) correctly accounts for the behaviour of the internal energy as a function of the bath temperature $T=\frac{1}{\beta}$.}
We also study the statistical distribution of the quantity $w$ defined as
\begin{equation}
    \label{eq:w}
    w(t^*)= \left(\frac{N}{E_{tot}N_c}{E_c}(t^*)-1\right)\sqrt{N_c}\,,
\end{equation}
for $t^*\gg 1$, to ensure that single trajectories are typical. In Fig.~\ref{fig:toda_distr} we show the distribution for some values of the total energy of the initial conditions.

\begin{figure}
    \centering
    \includegraphics[width=0.99\linewidth]{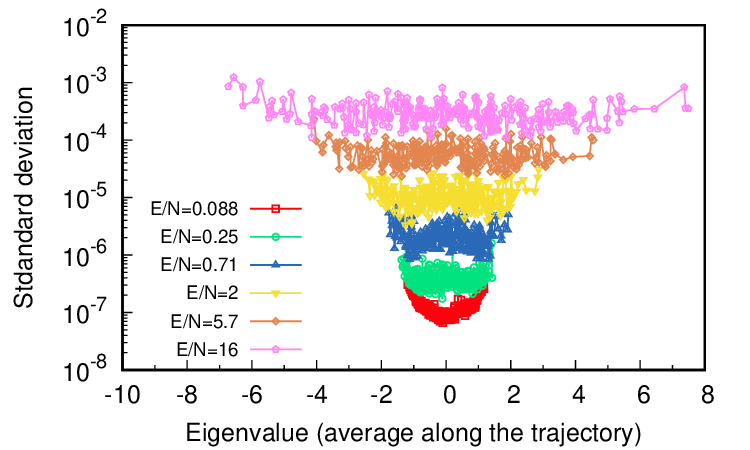}
    \caption{Standard deviations of the time averages on the trajectory of the eigenvalues of the Lax matrix, for some of the simulations in Fig.~\ref{fig:thermo}. Different symbols correspond to different energy levels of the Hamiltonian. The error on the conserved quantity is always smaller than $0.1\%$.}
    \label{fig:eigen}
\end{figure}

To ensure that the numerical results are not affected by numerical errors, we verify that each trajectory only explores a small volume around the torus defining the $N$ conserved quantities. To this end, we analyze the statistical distribution of each eigenvalue of the Lax matrix $L$. Fig.~\ref{fig:eigen} shows the standard deviation of this quantities computed for different values of the specific energy $\frac{E}{N}$. The standard deviation of the eigenvalues is always $\ll 1$, meaning that they are conserved with very accurate precision by our simulations.   
While the prediction for the potential energy derived from the canonical partition function Eq.~\eqref{eq:partition_function_potential} are in perfect agreement with simulation results, this might not be true when other observables are considered. For instance, when the specific heat is monitored, it shows a strong dependance on initial condition, implying that it cannot simply be predicted by the canonical equilibrium properties.  

\section{Conclusions}
\label{sec:conclusions}

In this work we have discussed the reasons why Statistical Mechanics successfully predicts the macroscopic properties of physical systems, even when the conditions for strong ergodicity do not hold. By revisiting analytical and computational results valid for integrable systems, we have provided support to Khinchin's point of view on the origin of typicality.
In the present paper we have focused on the deterministic  case. 
The problem is of great interest also for stochastic systems exhibiting ergodicity breaking, e.g. glassy dynamics, but we have not considered this class of models here. Indeed, for Markovian dynamics there are rigorous results allowing to establish the ergodic properties of the model, while the case of deterministic evolutions is much more involved.

In the context of linear chains, we have discussed the convergence  of the empirical distribution of the single-particle momentum $p_n(t)$ to a Maxwell-Boltzmann, even when the initial condition is far from equilibrium. The results presented in Sec.~\ref{sec:harmonic1} provide strong support to the idea of Khinchin about the definition of the thermalization property as characterizing an observable rather than a system. While there exists a reference frame (that of the eigenmodes) in which thermalization and equilibration will never occur, there are infinitely many others for which equilibration is a typical property. Moreover, we have shown that the distribution of infinite time averages of macroscopic observables sharply peaks around its equilibrium value for the majority of initial conditions even when localization occurs. The analysis presented in Sec.~\ref{sec:toda} on the Toda chain proved that the presence of such weak ergodicity is not a special feature of linear systems.

In our opinion, these results provide strong hints that the equilibrium statistical properties of physical systems are due to the large $N$ limit and to the choice of physically meaningful observables rather than to the properties of the microscopic dynamics. 

\section*{Data availability statement}
The code and data are available at~\cite{repositorycodes}.



\bibliography{biblio}

\end{document}